\documentclass[9pt,twocolumn,twoside]{pnas-new}

\templatetype{pnasresearcharticle} 
\nolinenumbers
\title{Complex 
pathways {and memory} in compressed corrugated sheets.}

\author[a]{Hadrien Bense}
\author[a,b]{Martin van Hecke}

\affil[a]{AMOLF, Science Park 104, 1098 XG Amsterdam, The Netherlands}
\affil[b]{Huygens-Kamerlingh Onnes Lab, Leiden University, PObox 9504, 2300 RA Leiden, The Netherlands}

\leadauthor{Bense}

\significancestatement{Describing the response of materials is central to physics, yet the complex response of frustrated media---featuring multi-step pathways which can be encoded by sequences of switching material bits---is poorly understood. Here we
introduce compression of corrugated elastic sheets to experimentally observe and manipulate these bits, their interactions, and their pathways. We
apply spatial gradients at the boundaries as a general technique to control the pathways. Besides mechanical memory, where the sheet remembers earlier extremal driving, we observe a pathway where interactions between material bits produce an elementary computation: counting to two.
Our work paves the way for controlling material bits in other systems, and suggests a tight link between computations and the complex response of frustrated matter.}

\authorcontributions{H.B. and M.v.H. conceived of the system and experiment, H.B. performed the experiments
and analysis, H.B. and M.v.H. wrote the manuscript.}
\authordeclaration{The authors declare no competing financial interests.}
\correspondingauthor{\textsuperscript{2}To whom correspondence should be addressed. E-mail: hbense.h@gmail.com}


\begin{abstract}
The nonlinear response of
driven complex materials---disordered magnets,
amorphous media, crumpled sheets---features intricate transition pathways where the system repeatedly hops between metastable states. 
Such pathways encode memory effects and may allow information processing---yet tools are lacking to experimentally observe and control these pathways, and their full breadth has not been explored.
Here we introduce compression of corrugated elastic sheets to precisely observe and manipulate {their full}, multi-step pathways, which are reproducible, robust,
and controlled by geometry.
We show how manipulation of the boundaries allows
to elicit multiple targeted pathways from a single sample. In all cases, each state in the pathway can be encoded by the binary state of material 'bits' called hysterons, and the strength of their interactions plays a crucial role. In particular,
as function of
increasing interaction strength, we observe Preisach pathways, expected in systems of independently switching hysterons, 'scrambled' pathways that evidence hitherto unexplored interactions between these material bits, and
'accumulator' pathways which leverage these interactions to perform an
elementary computation.
Our work opens a route to probe, manipulate and understand complex pathways, impacting
future applications in soft robotics and information processing in
materials.

\end{abstract}

\dates{This manuscript was compiled on \today}
\doi{\url{www.pnas.org/cgi/doi/10.1073/pnas.XXXXXXXXXX}}

\begin{document}
\nolinenumbers
\maketitle
\thispagestyle{firststyle}
\ifthenelse{\boolean{shortarticle}}{\ifthenelse{\boolean{singlecolumn}}{\abscontentformatted}{\abscontent}}{}

\dropcap{T}he response of complex media to external driving is intermittent, featuring smooth reversible episodes, associated with a single {(meta)stable state} 
of the system, punctuated by sharp irreversible steps between {states} 
that together form a multi-step pathway \cite{Nagel14,Wieker18,Adhikari2018,Regev19,Mungan21,Matan02,Lahini17}.
{These steps are typically hysteretic, and for several systems, such as amorphous media, can be associated with local rearrangements that act as 2-state degrees of freedom. The ensuing complex pathways are often modelled by collections of hysteretic, two-state elements called {\em hysterons} \cite{Preisach35}.}
These two-state hysteretic elements switch up and down between internal states $s=0$ and $s=1$ when a driving field $U$ passes through the upper and lower switching fields $U^+$ or $U^-$ (with $U^+ > U^-$);
the state of the hysteron for $U^- \le U \le U^+$ depends on its driving history.
One can think of these as 'material bits' \cite{,Serra-Garcia19,Pascall19,Reis21, Jules21} that collectively label the (meso)state of the physical system.
Properties such as memory are then determined by  the sequences of
bit switches as function of a global driving $U$, which can
be encoded in so-called transition graphs {(t-graphs)}, whose nodes represent the mesostates and edges their transitions \cite{Keim2019,Mungan2019}.

Collections of $n$ uncoupled  hysterons form the  Preisach model \cite{Preisach35}, which has been studied extensively in the context of complex hysteresis and memory effects.
The absence of coupling 
implies that hysteron $i$ changes state at switching fields $U_i^+$ and $U_i^-$, independent of the state of the other hysterons. As a result, the sequence of bit switches in response to sweeping $U$
is given by the ordering
of the $2n$ switching fields. This restricts the type of pathways that are possible, {with the t-graphs featuring a hierarchical structure of
loops within loops and
exhibiting}
return point memory (RPM), the widespread ability of complex systems to `remember their {extremal} driving', i.e., to return to a previous state when the driving revisits an {extremum} \cite{Barker83,Whitton52,Deutsch04, Shore93,Friedman05}.

\begin{figure*}
\centering	
\includegraphics{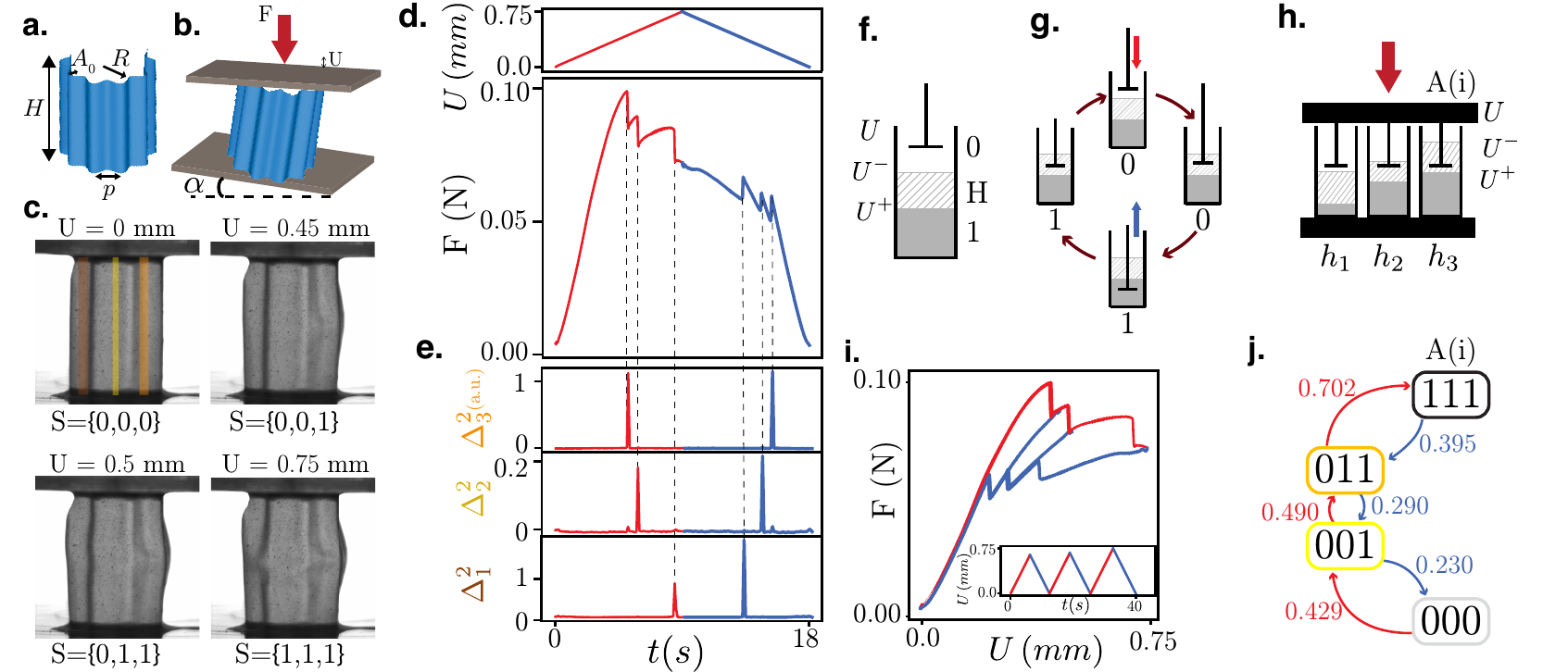}
\caption{\label{fig1}{\bfseries Robust pathways in a cyclically compressed corrugated sheet.}
\textbf{a}, Our samples are corrugated elastic cylindrical shells of height $H$, thickness $t$, radius of curvature $R$, and $N$ sinusoidal corrugations of pitch $p$ and amplitude $A_0$.
For sample A, shown here, $N=3$ and $\{H, t, R, p, A_0\} = \{35,0.2,1.0,8,3\} $ mm.
\textbf{b}, Compression $U$, force $F$ and bottom plate tilt angle $\alpha$.
\textbf{c}, Upon compression at $\alpha=0$ mrad, sample A reaches four different mesostates, associated with
sudden snapping of distinct regions $\Omega_i$ (colored strips, see Supplementary Movie).
\textbf{d}, The force $F$ exhibits three sharp jumps during compression (red) and decompression (blue).
\textbf{e}, Each force jump is associated with a sudden deformation, evidenced by spikes in
$\Delta^2_i$, the sum of the squared differences between subsequent digital images restricted to $\Omega_i$
(see Methods).
\textbf{f}, Schematic representation of a hysteron, its {two states (here grey corresponds to state 1, white to state 0, and dashed to the hysteretic range where the state is either 0 or 1 depending on the history)}, and the switching fields $U^\pm$, where $U^-<U<U^+$ is the hysteretic range.
\textbf{g}, Evolution of hysteron state during a compression cycle.
\textbf{h}, Our samples behave as collections of parallel hysterons with distinct thresholds.
\textbf{i}, Force-displacement curve corresponding to increasing compression cycles (inset). {The mechanical response of the system features connected hysteresis loops, and multiple pathways.}
\textbf{j}, The transition graph of sample A at $\alpha=0$ mrad contains
four states (nodes) labeled by the state of each hysteron.
Red (blue) arrows correspond to up (down) transitions at
(de)compression $U_c$ as indicated in mm.
}
	\end{figure*}

{However, interactions between hysterons can break the no-passing property that underlies RPM \cite{Shore93, Mungan21}. Recent simulations of models of interacting hysterons, as well as amorphous media, have presented examples for complex pathways and transition graphs featuring, e.g., avalanches, transient memories and multi-periodic orbits, which cannot be captured by models of non-interacting hysterons \cite{Nagel21, Paulsen21,Mungan21,VanHecke21}}.
{Unfortunately, distinguishing, observing and manipulating individual hysterons and their interactions is experimentally challenging for most complex systems. Moreover, we lack a conceptual framework that organizes the distinct impacts that hysteron interactions have on the phenomenology.} Hence {both} the connection between
hysteron models and experimentally observable pathways, as well as the relevance of hysteron interactions for driven complex media, remain unclear.

Here we introduce mechanical compression of
curved, corrugated elastic sheets to directly observe mechanical hysterons, their interactions and their concommitant non-trivial pathways (Fig.~\ref{fig1}). {We experimentally observe that the driving value where a given hysteron switches is modified by the states of the other hysterons, thus evidencing interactions between hysterons. To organize the resulting phenomenology, we distinguish between two characteristics of the pathways that are impacted by hysteron interactions. Most strikingly, interactions can modify the topology of the transition graph, and we identify the first three steps in a hierarchy of increasingly complex t-graphs and give concrete examples of each.  In addition, we show that even for a given t-graph topology, interactions can have a more subtle effect depending on the precise ordering of the switching fields. The strict hierarchy of t-graph topologies and the more subtle effects of the relative ordering are experimentally observable and testable, and provide a conceptual tool to organize the plethora of pathways observed in driven frustrated matter.} Together, our work
shows how hysteron interactions bring sequences of bit flips that encode forms of information processing within reach, creating new opportunities for soft robotics~\cite{Shepherd2011, Overvelde2015,Wehner2016,Coulais2018} and information processing in
materials~\cite{ McEvoy2015,Serra-Garcia19,Pascall19,Reis21}.

\section{Results and discussion}
\subsection*{Mechanical hysterons with robust pathways}
In our design, the corrugations lead to spatially localized instabilities upon compression which act as mechanical hysterons, the overall curvature prevents
global buckling of the sheet, and the open cylindrical structure allows to limit the number of grooves and facilitates both observation and manufacture.
Our experimental protocol involves sweeping the axial compression of a groovy sheet while filming the sample and measuring the compressive force $F$ (Fig.~\ref{fig1}a-e; see Methods and Supplementary Movies).
We observe that our samples exhibit sequences of well-defined steps, seen as
sharp jumps in the force $F$, and find that each event is associated with a localized (un)snapping event in a vertical groove, similar to those
seen in tape springs \cite{Bourgeois2012} (Fig.~\ref{fig1}c-e, Methods and Supplementary Movies). These transitions are hysteretic
and we observe that each groove can be in two distinct states---snapped and unsnapped---
so that each groove acts as a mechanical hysteron.
We refer to the hysteron transitions as 'up' (from the unsnapped to the snapped state) and 'down' (snapped to unsnapped), and denote the corresponding compression thresholds by the switching fields $U^+$ and $U^-$ (Fig.~\ref{fig1}f-h).

Repeated compression loops yield highly reproducible pathways with virtually identical force curves and sequences of hysteron-flipping, evidencing the irrelevance of creep, plasticity or aging (see Supplementary Information).
A wide range of groovy cylinders responds similarly to cyclical compression {(see Methods)}, and while we focus on systems with three grooves/hysterons, which is the minimal number for 
{ scrambled pathways (defined below)}, we have observed similar phenomena in larger systems (see Supplementary Information). As we will show below,
modifications of the sheet's shape and boundary conditions allow to geometrically tune the properties and interactions of their hysterons,
making this system a viable platform to study
reproducible, directly observable, and tunable pathways.



\begin{figure*}[t]
\centering	
\includegraphics
{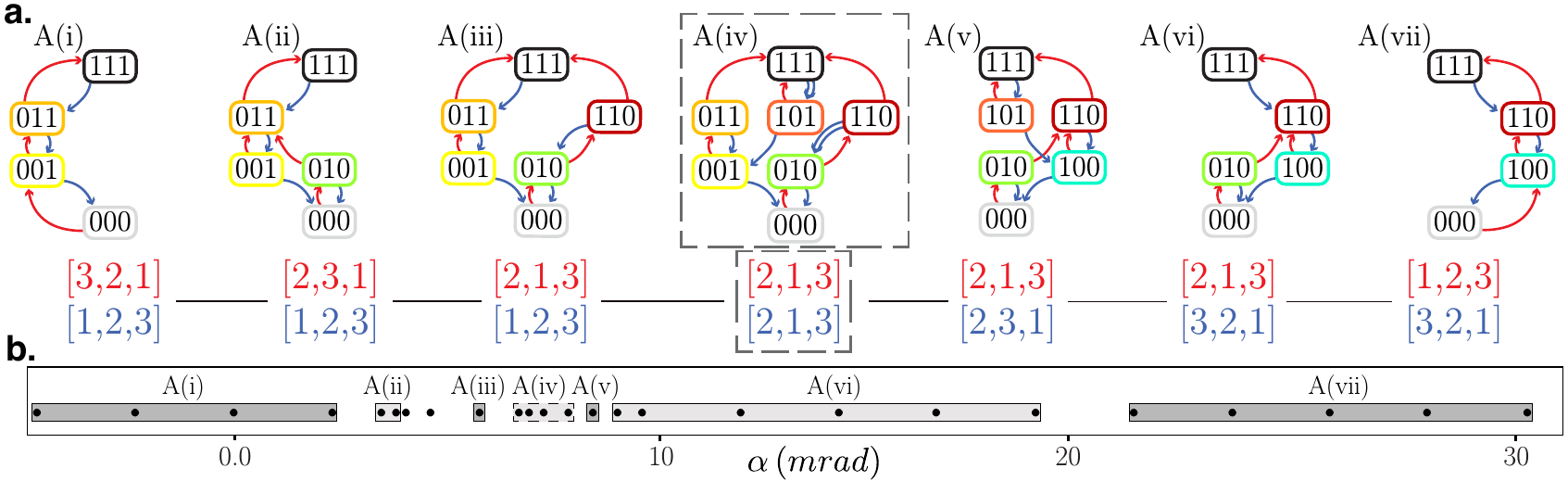}
	\caption{\label{fig2}{\bfseries Tunable transition graphs.}
\textbf{a}, Tilting the bottom boundary of sample A elicits six Preisach t-graphs and one scrambled t-graph (dashed box), where scrambled transitions are shown by double arrows.
The red (blue) lists indicate the order of the up (down) transitions of the main loop $P^+$ ($P^-$). The angle $\alpha$ increases monotonically from $A(i)$ to $A(vii)$. \textbf{b}, t-graph type as function of $\alpha$. 
The two unlabeled dots near $\alpha=5$ mrad refer to ambiguous cases where the critical switching fields $ U_1^+$ and $U_3^+$ are so close that the difference
between type (ii) and (iii) is not experimentally significant.}
	\end{figure*}

\subsection*{Pathways and transition graphs}
We map the full pathway by submitting the sample to a series of {well-chosen} compression/decompression cycles (Fig.~\ref{fig1}i).
Different driving cycles induce different pathways, which together form an intricate web of linked hysteresis loops \cite{Barker83,Whitton52} which connect distinct states.
These states  and their transitions can be collected into a {t-graph},
 a directed graph that captures the
response to {\em any} sequence of increases and decreases of the global driving field $U$ \cite{Regev19, Keim19, Keim2019, Mungan2019,Terzi20,Mungan21, Nagel21}.
To experimentally map the t-graph we systematically visit all states and determine all transitions, while tracking the state of each hysteron $s_i$, where $s_i = 1$ (0) refers to a snapped (unsnapped) state.
A state ${\bf S}$ is characterized by the hysteron states $ \{s_1,s_2,\dots\}$.
For each {collective} state---with the exception of the ground state {$\{0...0\}$ and saturated state $\{1...1\}$}---increased or decreased compression yields
'up' and 'down' transitions at critical switching fields $U_i^+({\bf S})$ and $U_i^-({\bf S})$.
To determine all transitions, we first determine the main loop,
the sequence of transitions that, under monotonic compression or decompression,
 connect the ground state  and saturated state  \cite{Mungan2019}.
We then check whether there are states for which there are undetermined
transitions, determine these, and whenever we obtain
a state that has not been previously visited we also determine its transitions,
repeating these operations until no new states are found.
Collecting all states and transitions we obtain the t-graph in which the nodes represent the mesostates, and the directed edges,
labeled by the values of their respective switching fields, the transitions
(Fig.~\ref{fig1}j). {We stress that in our systems, all states that we consider are mutually reachable, because there always is a specific driving protocol whereby any state is reachable from any other state \cite{Mungan2019}.}

For the simple case shown in Fig.~\ref{fig1}i-j, starting out
at the uncompressed state
$\{000\}$ and monitoring the force and images, we find that
continued compression
yields a sequence $\{000\}\!\rightarrow\!\{001\}\!\rightarrow\!\{011\}\!\rightarrow\!\{111\}$;
decompression starting at the saturated state
yields a sequence $\{111\}\!\rightarrow\!\{011\}\!\rightarrow\!\{001\}\!\rightarrow\!\{000\}$.
For this specific example, a simple compression/decompression cycle (Fig.~\ref{fig1}d) is enough to obtain the full set of transitions. A more complex protocol (inset of Fig.~\ref{fig1}i) yields a force response with three {linked} hysteresis loops (Fig.~\ref{fig1}i) that illustrates the different pathways the system can follow, and hence the importance of the loading history.
The corresponding t-graph also features three sub-loops embedded in the main loop and is spanned by
four nodes and six edges;
the material bits simply switch on and off when the driving is swept up and down.

\subsection*{Tunable pathways}
\begin{figure}[t]
\centering	
\includegraphics
{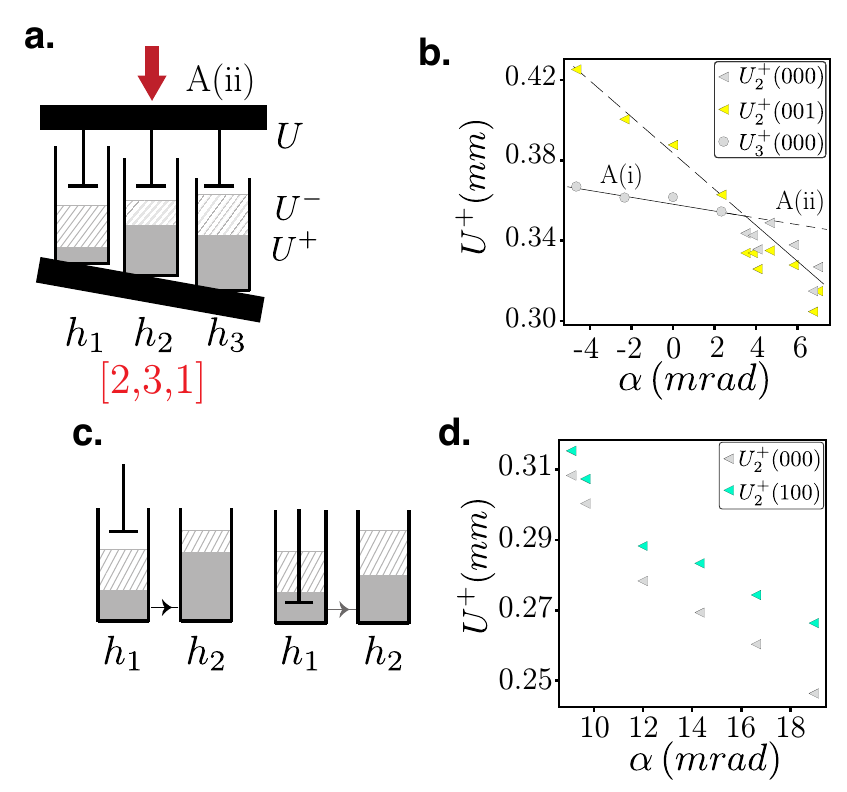}
	\caption{\label{fig3}{\bfseries Change of pathway and interactions.}
\textbf{a}, The boundary tilt
modifies the relation between local and global compression $U$, and can reorder the switching fields as indicated.
\textbf{b}, The switching fields vary smoothly with $\alpha$, and
the t-graph $A(i)$ is replaced by $A(ii)$ when $U^+_2(000)$ and $U^+_3(000)$ swap order.
\textbf{c}, Interactions cause the state of a given hysteron to modify the switching fields of another hysteron; here
hysteron 1 going $0\rightarrow 1$ increases $U_2^+$.
\textbf{d}, The measured difference between $U^+_2(000)$ and $U^+_2(100)$ indicates hysteron interactions as in panel (c).}
	\end{figure}

A wide variety of more complex t-graphs can be observed by tilting one of the boundaries of the sample (Fig.~\ref{fig1}b, Fig.~\ref{fig2}a-b, see Methods).
Due to the spatial separation of each hysteron, applying such global gradients in the driving modifies the relation between global and local compression, leading
to the smooth tuning of the switching fields $U_i^\pm$ (Fig.~\ref{fig3}a and Supplementary Information). As the relative order of the switching fields determines
{the order in which hysterons flip, tilting allows to visit different states and/or sequences, thus modifying the topology of the t-graphs.}
By sweeping $\alpha$, we observe seven distinct responses in sample A (Fig.~\ref{fig2}a). We characterize the order of the switching fields of the main loop by the corresponding sequence of hysteron flips,
$P^+$ and $P^-$ (Fig.~\ref{fig2}a).  For example, in the main loop of sample A in regime (ii), the second hysteron flips first ($\{000\}\!\rightarrow\!\{010\}$), followed by the third hysteron ($\{010\}\!\rightarrow\!\{011\}$) and finally the first one ($\{011\}\!\rightarrow\!\{111\}$), yielding $P^+ = [2,3,1]$. Similarly, during decompression, the first hysteron unflips first ($\{111\}\!\rightarrow\!\{011\}$), then the second ($\{011\}\!\rightarrow\!\{001\}$) and finally hysteron number three ($\{001\}\!\rightarrow\!\{000\}$), leading to $P^-=[1,2,3]$.
Many of the t-graphs exist on a large angle span, while others can only be observed for a limited range of tilt angles $\alpha$ (Fig.~\ref{fig2}b). We note that during the tilting process, two switching fields can become extremely close, such that their respective (un)snapping events become indistinguishable (e.g., in {Fig.~\ref{fig2}b} the two data points between A(ii) and A(iii) refer to such a case). This degeneracy may cause an avalanche. However, we can distinguish degeneracy-driven avalanches from true avalanches by their response to changes in the tilt angle: degeneracy-driven avalanches quickly disappear when the angle is modified, while true avalanches persist over a significant range of $\alpha$.\\
\indent We note that in most cases, the difference between  neighboring t-graphs is  associated with a single permutation of the upper or lower transitions; for example compare graph $A(i)$, where $P^+=[3,2,1]$  to $A(ii)$ where $P^+ = [2,3,1]$ corresponding to a swap of the snapping order of hysterons $2$ and $3$. Such permutation is consistent with the notion that tilting smoothly tunes the switching fields. Plots of the switching fields directly evidence their smooth variation
with the tilt angle $\alpha$, and show that the
ranges of $\alpha$ where each t-graph occurs are consistent with the crossing of two switching fields;
for example, the change from graph $A(i)$ to $A(ii)$ corresponds to the crossing of $U_2^+$ and $U_3^+$ (Fig.~\ref{fig3}b and Supplementary Information).
We have obtained a similar range of t-graphs for a sample C which contains four hysterons (see Supplementary Information).
We conclude that global driving gradients are a powerful tool to systematically elicit a multitude of t-graphs from a single sample.


	\begin{figure*}[t]
	\centering	
\includegraphics{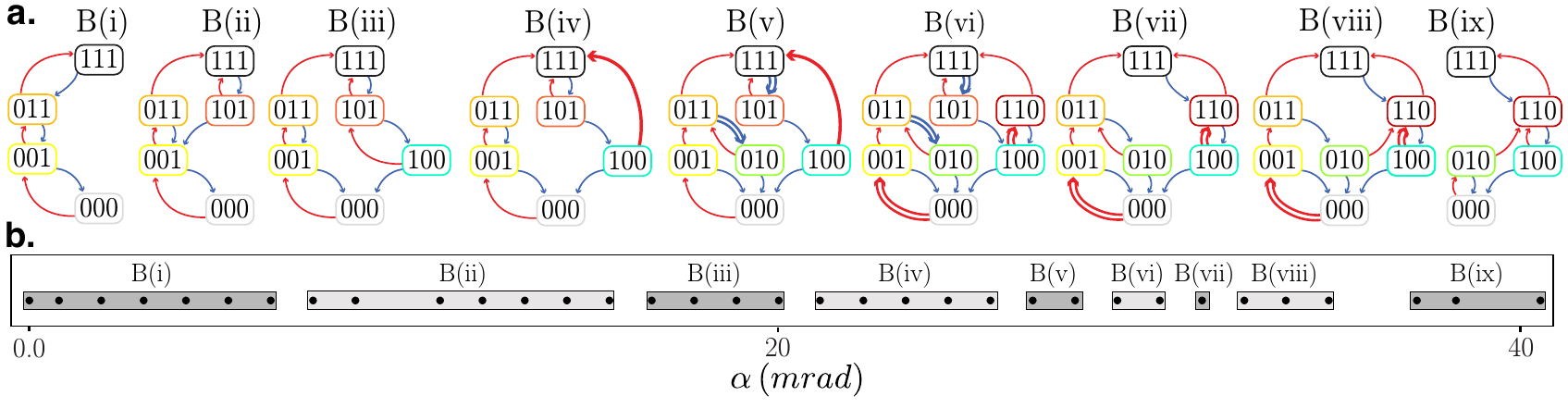}
	\caption{\label{fig4}{\bfseries t-graphs of sample B.}
\textbf{a},Tilting sample B ($\{H, t, R, p, A_0\} = \{20,0.2,2.0,10,3\} )$ mm and $N=3$ corrugations) elicits nine distinct t-graphs.
%
\textbf{b}, t-graph type as a function of $\alpha$.
}
	\end{figure*}


\subsection*{{Classification: Preisach and scrambled t-graphs}}
{Different pathways can be classified according to the topology of their t-graphs, and in this paper we introduce the first three steps of complex pathways that arise when interactions become increasingly important.}
A close inspection of the topology of the t-graphs of sample $A$ reveals {the first }
two distinct classes of graphs: Preisach graphs $A(i)$, $(ii)$, $(iii)$, $(v)$, $(vi)$ and $(vii)$ and scrambled graphs $A(iv)$.

{To understand this distinction, consider the relation between the flipping of the individual hysterons and the state transitions.
For $n$ hysterons with states $s_i$, forming a collective state $\bf S$,
the up (resp. down) transitions are set by the hysteron in state $s=0$ (resp. $s=1$) which has the lowest up (resp. highest down) switching field:
\begin{eqnarray}
U^+({\bf S}) = \min_{i_0} u_{i_0}^+({\bf S})~,\\
U^-({\bf S}) = \max_{i_i} u_{i_1}^-({\bf S})~.
\end{eqnarray}
Here $i_0$ (resp. $i_1$) runs over the hysterons that are in state 0 (resp. 1),
$U^{\pm}({\bf S})$ are the global switching fields for the collective states ${\bf S}$,
and $u_i^\pm$ are the state-dependent switching fields of individual hysterons.}

In absence of hysteron interactions, the switching fields of all hysterons are independent of the state of the other hysterons ($u^\pm_i({\bf S}) = u^\pm_i$), and the topology of the t-graph is thus fully determined by the
ordering of the up and down switching fields of individual hysterons.
We refer to t-graphs whose topology is consistent with a unique ordering of the upper and lower switching fields of the hysterons, and which do not contain avalanches,
as Preisach graphs (recall that collections of non-interacting hysterons are referred to as  the  Preisach model
 \cite{Terzi20}).
We note that
the topology of a Preisach graph is fully determined by the order of the transitions of the main loop, encoded in $P^+$ and $P^-$. We stress here that t-graphs with a
Preisach topology do not require the strict absence of interactions, but only that interactions do not introduce differences between the ordering of the transitions in the main loop and the rest of the t-graph.

Strikingly, we also observe a {\em scrambled} graph which features pairs of transitions that are not consistent with a unique, state-independent ordering of the switching fields
---we call these transitions scrambled.
In particular, t-graph $A(iv)$ contains the pair of transitions  $\{111\} \rightarrow \{101\} $ and  $\{110\} \rightarrow \{010\}$, which {is} not compatible with a unique, state-independent ordering of the switching fields $U^-_2>U^-_1$; rather these transitions  imply that
$U^-_2(111) >U^-_1(111)$ and
$U^-_2(110) < U^-_1(110)$ {respectively} (Fig.~\ref{fig2}a). Such state-dependent ordering of the switching fields is incompatible with a Preisach graph and 
demonstrates
a dependence of the switching field of one hysteron on the state of another hysteron, thus directly evidencing hysteron interactions.
{We note here that scrambled pairs of transitions have recently been observed in numerical simulations, e.g., the scrambled pair of  transitions {$\{000\}\! \rightarrow\!\{100\}$} and
$\{001\}\!\rightarrow\!\{011\}$ in Fig.~5b of \cite{Nagel21}, and the scrambled pair of
transitions $\{1101\}\!\rightarrow\!\{1100\}$ and
$\{1111\}\!\rightarrow\!\{0111\}$ in Fig.~2 of \cite{Paulsen21}; see also \cite{VanHecke21}.}
Intuitively, scrambling implies that bit-flip sequences depend on the starting state, allowing for a far larger space of potential t-graphs than Preisach graphs.
Additional evidence of interactions can be obtained
from the state dependence of the switching fields;
for example we observe a systematic difference of $U_2^+(000)$ and $U_2^+(100)$  (Fig.~\ref{fig3}c-d, see Supplementary Information). {We note that the sign of the interactions is not constant throughout the samples---ferromagnetic and antiferromagnetic generically both occur, and different pairs of hysterons can feature different signs of interactions. Moreover, interactions are not reciprocal, and the sign and strength of interactions for the upper and lower switching fields may also be different.}
We stress here that the interaction strength necessary to obtain scrambling is directly proportional
to the differences between the switching fields of different hysterons. Hence, as tilting allows to make these differences arbitrarily small near crossings, our strategy is eminently suited to observe non-Preisach behavior, even if hysteron interactions are weak.
We conclude that hysteron interactions can yield novel types of pathways and t-graphs.

\subsection*{Strong interactions}
{To further study the effect of interactions between hysterons we require samples with stronger coupling.} To create these, we note that for typical parameters and $R \gtrsim 5$ cm, the (un)snapping of one hysteron triggers the (un)snapping of all hysterons, which we interpret as strong interactions, while for smaller $R$ the snapping events occur in sequence. {Indeed, radius of curvature serves as} a proxy for the strength of interactions---the larger the radius, the stronger the interactions. We thus introduce sample B with $R=2$ cm (the radius of curvature for sample A was 1 cm) and three grooves. We have verified that,
in comparison to sample A, the interactions as measured by the dependence of the switching fields on the state are indeed stronger (see Supplementary Information).
We orient the sample and apply a shim to the boundaries
so that the crossings of the switching fields as function of $\alpha$ are optimally separated.

{We find that, as a function of the tilt angle $\alpha$, sample B yields nine distinct t-graphs, occurring on well-separated ranges of $\alpha$ (Fig.~\ref{fig4}). We distinguish a number of distinct features. First, t-graphs $B(i$), $B(ii)$, $B(iii)$ and $B(ix)$ are all Preisach graphs. Second, t-graphs $B(iv)$ and $B(v)$ feature an avalanche $\{100\}\rightarrow\{111\}$
over a significant range of tilt angles $\alpha$. While we cannot rule out that this avalanche is caused by inertial effects or degeneracies, we note the avalanche occurs over a significant range of $\alpha$, and that hysteron interactions can also cause such avalanches; if $U_3^+(110)<U_2^+(100)< U_3^+(100)$, state $\{100\}$ transitions to $\{110\}$ at $U=U_2^+(100)$, and since state $\{110\}$ is unstable at this value of $U$, it transitions to a stable state $\{111\}$. Third, t-graphs $B(v)$, $B(vi)$, $B(vii)$ and $B(viii)$
all contain scrambled transitions---see Table~\ref{tableScramble} for the pairs of scrambled transitions. Hence, manipulating the overall geometry of our corrugated sheets allows to increase the magnitude of interactions to obtain a variety of robust, non-Preisach pathways.}

\subsection*{{Breaking of l-RPM and accumulator pathway}}
{To further classify the t-graphs topologies, we consider the
recent definition of \emph{loop-RPM} (l-RPM) \cite{Terzi20,Mungan2019}. l-RPM requires that all loops within the t-graph are absorbing. In essence, a loop is defined by a pair of nodes $S_m$ and $S_M$, where the system evolves from $S_m$ to $S_M$ (and vice-versa) by a series of up (down) transitions, and the intermediate states are defined as the up (down) boundar{ies};  l-RPM then requires states $S_m$ ($S_M$) to be accessible by a sequence of down (up) transitions from any up (down) boundary.
All t-graphs of sample A, and all t-graphs of sample B---with the exception of $B(viii)$, which we will discuss in detail below---are consistent with l-RPM. }

{ We now focus on sample B in regime $(viii)$.
To see that its t-graph violates l-RPM, consider the loop with $S_m=\{000\}$ and $S_M=\{011\}$. State $\{010\}$ is then part of its down-boundary, but starting from this state and increasing the driving, never reaches $S_M=\{011\}$: hence this 
{loop}
is not absorbing, and the t-graph violates l-RPM.
Hence, in this sample hysteron interactions are sufficiently strong to observe a next step in the hierarchy of t-graph complexities, {that yields t-graphs with topologies that violate l-RPM}.}

We stress here that it is known that ferromagnetic interactions, where one hysteron switching from zero to one promotes the switching of others from zero to one and vice versa, preserve l-RPM\cite{Shore93}. {Hence our observation of weak and strong RPM breaking indicate the presence of anti-ferromagnetic interactions in our system, which clearly can cause a wide range of t-graphs \cite{Regev19, Nagel21, Paulsen21, VanHecke21}}.

\begin{figure}[t]	\centering	
\includegraphics{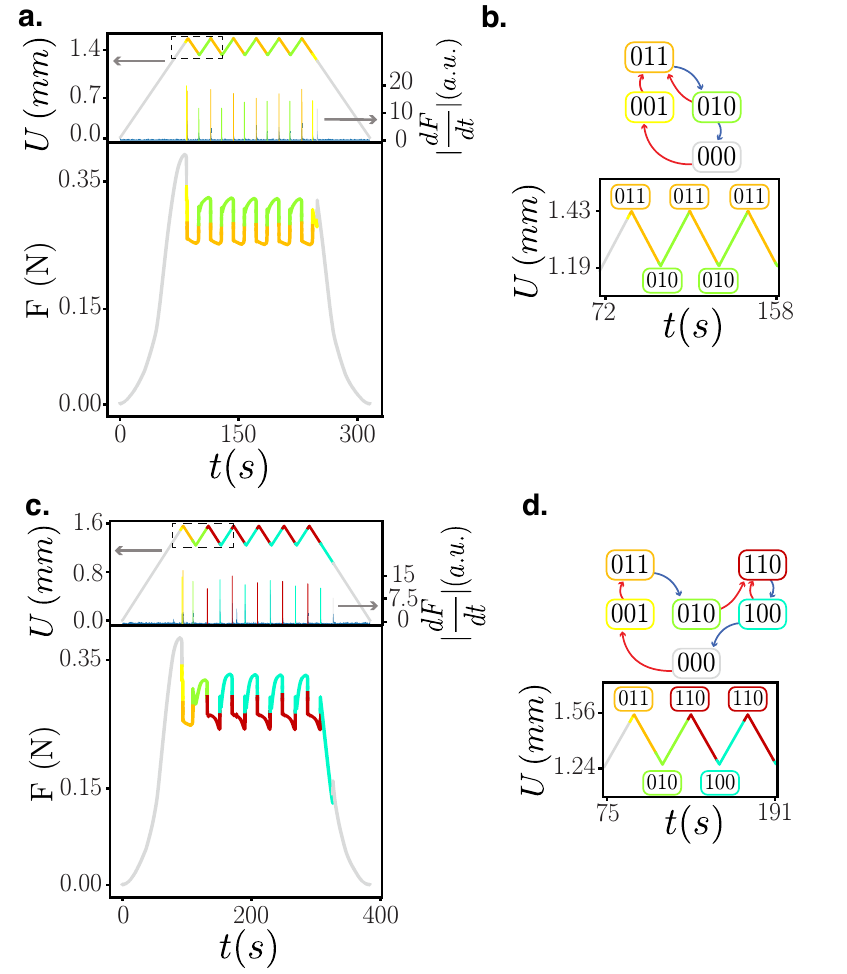}
	\caption{\label{fig6}{\bfseries Accumulator behavior.}
\textbf{a}, Repeated compression cycles yield repeated loops between the same extremal states (indicated by color) in regime $B(vii)$, consistent with RPM. Peaks of $|\frac{dF}{dt}(t)|$ indicate transitions.
\textbf{b}, Corresponding pathway. \textbf{c}, Repeated compression cycles in regime $B(viii)$ evidence `accumulator' behavior which violates RPM.
\textbf{d}, Corresponding pathway where the first and subsequent extremal states are different. }
	\end{figure}

To understand the new qualitative feature corresponding to this specific case
we compare the response to cyclic driving in regime $(vii)$, where the t-graphs {has} l-RPM, and regime $(viii)$, focussing on the loop that connects states $S_m=\{000\}$ and
$S_M=\{011\}$ (Fig.~\ref{fig6}a-d). {For cyclic driving, specified by $U_m$ and $U_M$, and a specific starting state $S$, the system must reach a periodic orbit since there are a finite number of states \cite{Mungan2019}. The system may require $\tau$ cycles before reaching its final orbit, and this orbit may be subharmonic with perioditicity $T$---simple cycles or absorbing states correspond to $T=1$ \cite{Mungan2019,Nagel21,Paulsen21}.} In regime $(vii)$, once the state $\{011\}$ is reached, cycling $U$ between $U_m (>U^-_2(010))$ and $U_M (<U^+_1(011)) $ the system follows the same $\{011\}$-$\{010\}$ loop repeatedly (Fig.~\ref{fig6}a,b).
In contrast, applying a similar driving to the sample for
$\alpha$  in regime $(viii)$,
yields state $\{011\}$ at first maximal driving, but the second and subsequent maxima produce state $\{110\}$ (Fig.~\ref{fig6}c,d). Hence, the transition $\{010\} \rightarrow \{110\} $ "erases" the memory of the $\{011\}$ state and brings the system to a new sub-loop. {The system thus reaches a simple periodic orbit ($T=1$) after a training of $\tau=2$ driving cycles. We refer to orbits where $\tau>1$ and $T=1$ as accumulators: during a transient of $\tau$ cycles, the system visits unique states from which the number of driving cycles can be deduced; after more driving cycles, the system visits the same orbit, which encodes that the number of driving cycles is larger or equal than $\tau$. Such orbits can be seen as a concrete realization of a 'long training time' pathway with $\tau =2$, as presented numerically in \cite{Nagel21}. This behavior, where the sample 'remembers' how often it is driven to a certain maximum is different from classical forms of memory \cite{Keim19}, and we suggest that t-graphs containing accumulator orbits  may underpin this behaviour, although we note that here we specifically focus on the subcase where $T=1$. We stress that {the observed} accumulator behaviour can also be seen as an elementary form of information processing: counting to two. From this standpoint, we suggest that our observation of accumulator behavior is a first step towards the realization of systems with complex t-graphs that encode information processing.} We {moreover} stress that this accumulator behavior \cite{Dehaene2012} can be observed on a robust range of tilt angles (see Fig.~\ref{fig4}b). {Finally we note that in our system, where all states are mutually reachable, the existence of an accumulator orbit necessitates the breaking of l-RPM \cite{Mungan2019} \footnote{{If not all states are mutually reachable, one can construct more complex t-graphs that contain multiple maximal loops that cannot be parts of a single main loop. Orbits between multiple maximal loops can then feature arbitrary long transients and subharmonic behavior, even when the system has l-RPM. For more details see \cite{Mungan2019}.}}.}

{We conclude that our system allows to observe three classes of topologically distinct t-graphs. The first class consist of Preisach t-graphs which are topologically equivalent to those of the Preisach model. The second class consist of t-graphs that contain scrambled transitions, yet satisfy l-RPM. The third step consist of t-graphs that violate l-RPM, and in our specific example the t-graph encodes {\em accumulator} behavior, which may underpin transient memories \cite{Nagel21}. This classification, which is experimentally accessible, organizes the impact of hysteron interactions on the pathways.
}

\begin{table}[ht]
\centering
\begin{tabular}{p{0.2\linewidth}p{0.3\linewidth}p{0.3\linewidth}}
    \hline
t-graph & $t_1$ & $t_2$ \\
 \hline
$B(v)$ &$\{111\}\rightarrow \{101\}$ & $\{011\} \rightarrow \{010\}$\\ \hline
$B(vi)$ &$\{111\}\rightarrow \{101\}$ & $\{011\} \rightarrow \{010\}$\\ \hline
$B(vi)$ &$\{000\}\rightarrow \{001\}$ & $\{100\} \rightarrow \{110\}$\\ \hline
$B(vii)$ &$\{000\}\rightarrow \{001\}$ & $\{100\} \rightarrow \{110\}$\\ \hline
$B(viii)$ &$\{000\}\rightarrow \{001\}$ & $\{100\} \rightarrow \{110\}$\\ \hline
    \end{tabular}
    \caption{\label{tableScramble} Pairs of scrambled transitions $(t_1,t_2)$ in Sample B.}

\end{table}

{
\subsection*{Ordering of the switching fields} 
The topological properties characterize that the system can reach certain states, but {do not restrict the} corresponding values of the driving. However, as we demonstrate  below, the precise ordering of the switching fields may impact the pathways, even if it does not impact the t-graphs topology. Hence the ordering of the switching fields provides a secondary characterization of pathways.}

{An important example of a property that involves a precise statement on the values of the switching fields is ordinary return point memory (RPM), which requires that
when the driving strength revisits a previous {\em extremal value},  the system revisits a previous extremal state \cite{Terzi20,Keim19,Barker83,Shore93,Whitton52}.} RPM implies l-RPM, as all extremal states are absorbing states when RPM is valid \cite{Mungan2019}. {The reverse however is not true: RPM implies conditions on the relative ordering of the switching fields beyond those captured by the topology of the t-graph}: a system can be l-RPM without strictly satisfying RPM.

\begin{figure}[t]
\includegraphics[scale=1]{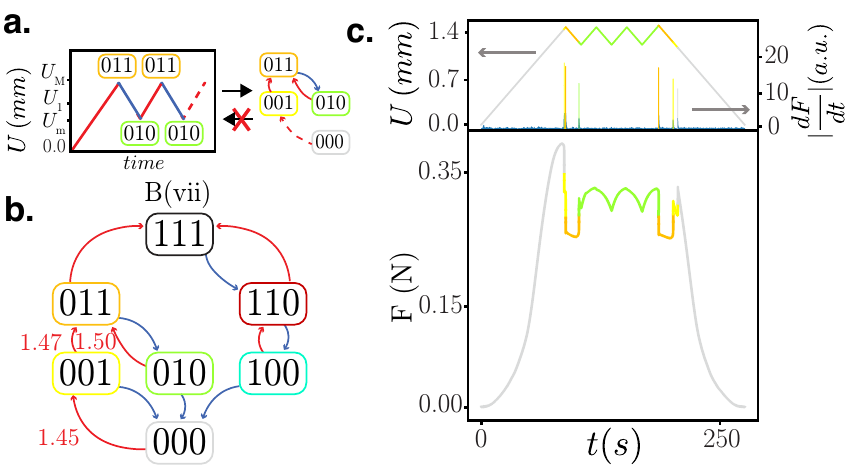}
\caption{\label{fig5}{\bfseries Breaking RPM.}
\textbf{a}, While a strict RPM (left part) implies l-RPM (right part), l-RPM does not imply strict RPM. \textbf{b}, $C(vii)$ is a scrambled t-graph with l-RPM. However, as  $U_2^+(001)< U_3^+(010)$ (indicated in mm), this pathway can violate RPM and satisfy l-RPM. \textbf{c}, Cycles between $U_M$, with $U_2^+(001)<U_M< U_3^+(010)$ (we take $U_M=1.476\,$mm) and $U_m>U_2^-(010)$ (we take $U_m=1.210\,$mm) leave the system in state $\{010\}$, despite $U_M$ being larger than the critical up transition value of state $\{001\}$, thus breaking strict RPM. In the last cycle we increase $U$ beyond $U_3^+(010)=1.50\,$mm and reach state $\{011\}$. Peaks of $|\frac{dF}{dt}(t)|$ indicate transitions}
\end{figure}

{To illustrate the importance of the ordering of the switching fields, consider
a scenario where state $\{011\}$ can be reached by up transitions from either $\{001\}$ or $\{010\}$ (Fig.~\ref{fig5}a).
First increase $U$ to a value $U_M$ so that state $\{011\}$ is reached, then decrease $U$ to $U_m$ so that state $\{010\}$ is reached. Now consider the response to increasing $U$
back up again to $U_M$.
If the system satisfies RPM, it must then revisit state $\{011\}$ (and not remain stuck at $\{010\}$), which implies the following condition on the ordering of the switching fields: $U_2^+(001) \ge U_3^+(010)$. For a Preisach system, the sequence of transitions on the main loop implies that $U_2^+> U_3^+$, so that this condition is satisfied. However, in the presence of hysteron interactions, the sequence of transitions on the main loop only implies that $U_2^+(000)> U_3^+(000)$ and $U_2^+(001)> U_3^+(000)$, but does not imply that $U_2^+(001) \ge U_3^+(010)$; indeed, for the same t-graph topology, RPM can be satisfied or violated, depending on the ordering of the switching fields.}

{A specific example of such a t-graph that satisfies l-RPM but violates RPM
is sample B in regime $(vii)$ (Fig.~\ref{fig5}).
Here, $U_2^+(001)\approx 1.47$ mm and
$ U_3^+(010) \approx 1.50$ mm. Hence, when we
cycle the system (starting from $\{010\}$) between $U^M=1.476\,$mm and $U^m=1.210\,$mm, the system reaches state $\{011\}$ in the first cycle but then
remains stuck in the $\{010\}$ state; reaching $\{011\}$ again requires raising
$U$ beyond  $U_3^+(010)$ (Fig. \ref{fig5}c).
Hence, sample $B(vii)$ does not satisfy ordinary RPM, because the condition
$U_2^+(001) \ge U_3^+(010)$ is violated.}

{This is an example of the violation of the 'No-Passing' (NP) property. The topology of the t-graph (Fig.~\ref{fig5}b) implies the following ordering {green}{of the states}: $\{000\} \prec
\{001\} \prec \{011\}$ and $\{000\} \prec
\{010\} \prec \{011\}$ \cite{Mungan2019}.
If NP were to hold, two orbits starting out at different states, e.g., $\{000\}$ and $\{010\}$ must preserve this ordering (equalities allowed) when they occur at the same driving.
However, starting from these states and ramping the driving to, e.g,  $U^M=1.476$ mm,
the orbit $\{000\}\!\rightarrow\! \{001\}\!\rightarrow\! \{011\}$ 'passes'
the other orbit that remains stuck at $\{010\}$.
}

{This example of NP and RPM violation, which does not affect the t-graph's topology, exemplifies the role of the precise ordering of the switching fields.
We note that similar effects can also occur in more complex t-graphs; for example, in the accumulator t-graph, some cyclic driving protocols can remain stuck at $\{011\}$ only when $U_2^+(001) < U_1^+(010)$.
We finally stress
that while both topology changes, such as scrambling, and secondary effects, such as RPM breaking,
relate to ordering of switching fields, they are distinct. E.g., scrambling requires specific differences in the ordering of two switching fields in two states, and changes the topology of the t-graph; secondary
ordering effects do not impact the topology of the t-graph, and can be associated with single inequalities. Hence, by clearly distinguishing topological and non-topological impact of hysteron interactions, transition graphs and pathways can precisely be characterized and classified.}

\section{Concluding remarks}
{We presented a system with experimentally accessible mechanical hysterons which allows to study emergent complex pathways. Our work shows that hysteron interactions yield a plethora of t-graphs, properties, and distinct flavors of return point memory.
In particular, our study elucidates that a complete classification of pathways combines a characterization of the t-graphs topology, as well as aspects of the exact ordering of the switching fields. For the topology, we have observed the first three steps in a hierarchy of increasing complexity as Preisach t-graphs, scrambled t-graphs, and t-graphs that break l-RPM. We discussed how the accumulator t-graph is a concrete example of transient memory, and may underpin recent observations of transient memory in models of interacting hysterons \cite{Nagel21}. We demonstrated the secondary effects due to state dependent switching fields by the specific example in which strict RPM is broken, irrespective of the t-graphs topology{.
We introduced spatial gradients as a general experimental strategy to modify the pathways, which moreover shows that specific behaviors are not a only property of specific samples but can be tuned geometrically via their boundaries. Finally, we stress
that the t-graph hierarchy and the secondary effects of state-dependent switching fields are experimentally observable and testable.}}

Our work emphasizes the proliferation of complex t-graphs in complex media. Studies that explore t-graphs in such frustrated systems have just started to emerge and  {mostly} focus on numerics 
\cite{Nagel21, Paulsen21,Mungan21,VanHecke21}, and we hope our {work motivates} further studies by experiments as well {see also
\cite{Jules21,Lahini21}}. At present, we have little information about the statistics of different classes of t-graphs, the levels of complexity that can be reached and observed, and their relations to physical properties of the underlying system, all of which provide fertile ground for further study.

Our work further suggests to investigate the utility of complex t-graphs in rationally designed metamaterials. Finding strategies to arbitrarily control hysteron properties and their interactions, beyond the boundary control method introduced here, may
open up a large design space for the rational design of pathways and t-graphs.
Formally, the t-graphs have the same structure as the directed graphs that encode computations by finite state machines \cite{Sakavoritch09}.
Hence, we suggest that a fruitful perspective on t-graphs in complex matter starts from their information processing capabilities. We note that
while our systems are purely elastic and thus microscopically reversible, one imagines that material plasticity will lead to evolution of such pathways, which perhaps can be used to train materials to exhibit targeted pathways. Together, such control, design and learning strategies can be explored, in particular in systems with many hysterons, to achieve mechanical systems which, in response to external driving, process complex information.

\matmethods{


\subsection*{Sample fabrication and experimental protocol}
The fabrication of corrugated sheets starts by spin coating a liquid mixture of a two components silicone elastomer (Zhermack Elite double 32 Fast, Young's modulus $E\approx 1$ MPa, Poisson's ratio $\nu \approx 0.5$) on a surface with sinusoidal corrugations with pitch $p$ and amplitude $A_0$. Rotation is maintained until complete curing of the polymer ($\approx 20\,$min). The sheet is then peeled and rolled in an open cylinder; top and bottom ends are dipped in a liquid layer of the same polymer mixture to set the cylindrical shape and fix the boundary conditions. The resulting sample is characterized by its height $H$,  thickness $t$, radius of curvature $R$, pitch $p$ and
number of corrugations $N$, and amplitude $A_0$. {We have observed similar local and sequential snap-through behavior in over ten samples, with the only limitation appearing to be that the corrugation amplitude $A_0$ is not too small and $R$ is not too large--a natural scale to compare these to is the pitch $p$.} Paint is splattered on the samples to enhance contrast and ease visualization.

The mechanical response of our samples is probed in a uniaxial testing device (Instron 3366) which controls the axial compression $U$ better than $10\,\mu$m; we use a $5$N sensor which accurately measures the force down to $10$ mN with an accuracy of $10^{- 4}\,$N.
We define $U=0$ where the force during compression reaches the small value $F(U_0) = 20\,$mN.
We use compression speeds of $1\,$mm/min and have checked that further lowering
the compression speed by an order of magnitude does not affect the phenomenology,
thus ensuring we operate in the quasistatic regime.
We focus on the compression range (strain less than $5\%$) where grooves can snap but where no additional instabilities are observed.

We image the deformation of the groovy sheet during compression at a frame rate of $3$Hz or faster, using a CCD camera (Basler acA2040-90um) mounted with a $50\,$mm objective. We calculate the mean squared differences in each local region $\Omega_i$ of the normalized digital image as: $ \sum_{k,l \in \Omega_i}\Delta_i^2 {k,l} :=  (A_{k,l}^{t+\Delta t} - A_{k,l}^t )^2$, where $k$ and $l$ label the pixels, $t$ is time, and $\Delta t$ the time interval ($\Delta t = 100\,$ ms in Fig. 1e). Each region $\Omega_i$ targets a part of a single groove, chosen such that events in neighbouring regions do not create secondary peaks.

The sample rests on a Thorlabs tilt stage that allows to control the tilt angle $\alpha$ with an accuracy of $3.10^{-5}\,$rad. We incrementally change $\alpha$ with steps ranging from $3.10^{-4}\,$rad to $2.10^{-3}\,$rad, and for each tilt angle $\alpha$ measure the full t-graph and the mechanical response.
All transition graphs presented in this paper were obtained multiple times over the course of several weeks, and all angles were visited several times to ensure a good reproducibility (see Supplementary Fig.~S1c,d). By exploring the reproducibility of the boundary between different t-graphs, when two switching fields are essentially degenerate, we estimate our accuracy of the boundaries to be better than $\pm 2.10^{-4}$ rad.


To determine the switching fields, each transition is probed between two and four times, and we report mean switching fields which have a standard deviation typically smaller than the symbol-size. We estimate small viscous relaxation effects to affect the switching field by at most $4\%$ (Supplementary Fig. S1b), thus requiring larger differences to evidence interactions.

}

\showmatmethods{} 

\acknow{We gratefully acknowledge discussions with C. Coulais, M. Mungan, E. Verhagen, N. Keim and J. Paulsen. {In addition, we thank M. Mungan for pointing out connections to the no-passing property. Finally we thank M. Teunisse for insightful discussions about $\tau$ and $T$ definitions, and their connection with accumulator behavior.}}

\showacknow{} 

\bibliography{pnas-sample}

\end{document}